# Deriving real delay time statistics from the complex delay time statistics in weakly disordered optical media

Prabhakar Pradhan, Peeyush Sahay, and Huda M. Almabadi

*BioNanoPhotonics Research Laboratory*
*Department of Physics and Materials Science*
*University of Memphis*
*Memphis, TN 38152*

Considering the complex reflection amplitude $R=r^{1/2}\exp(i\theta)$ of a light wave, real delay time $\tau_r$ (i.e., sojourn or Wigner delay time), which is the energy derivative of the real phase ($\tau_r = d\theta/cdk$), and complex delay time $\tau_i$, which is the energy derivative of the reflection coefficient ($\tau_i = d\theta_i/cdk$, $|R|=r^{1/2}=\exp(-\theta_i)$), have the same statistical form and a mirror image with a shift in time in weak disorder and short length regime. Real delay time statistics obtained from the reflection coefficient can be attributed to the strong correlation between the reflection coefficient and its phase in this regime.

## 1. Introduction

Complex reflection amplitude $R=r^{1/2}\exp(i\theta)$ has two parts: the absolute value $r^{1/2}$ and its associated phase $\theta$. In general, the reflection coefficient, or intensity $RR^*$ ($=r$), has no phase information. To obtain phase information, information about the complex reflection amplitude is required [1]. In a real case scenario, interference experiments must be performed simultaneously with respect to the phase of a reference signal in order to keep track of the phase of the reflected wave in a systematic manner [2]. However, either keeping track of the complex amplitude information analytically or generating phase information by using the interference method is a tedious process. In this context, we asked how to obtain phase information from reflection intensity in a case where the apparent phase information is generally missing. If such case were to present itself, we further asked how to obtain information of any other physical parameters that depend on phase or its derivatives, in particular delay time (or sojourn time) which depends on the derivative of the phase with respect to the energy or wave vector. Properties of delay time have been studied recently for both electronics and optics [3-15]. In this paper, we will mainly consider delay time studies of optical systems. In general, through scattering from a deterministic potential, phase and amplitude information can be determined by directly solving the optical wave equation using the known potential function [1]. Consider a wave reflected from a disordered optical potential, or refractive index medium. In most parameter regimes with weak disorder and large length, it has been shown that the phase and amplitude of a reflected wave are statistically independent and that the phase has uniform distribution statistics [16]. Therefore, phase and amplitude of the reflection are two independent statistical variables for weak disorder and large length of the sample.

It was recently shown that the phase and amplitude of a complex reflection amplitude are strongly correlated for a weakly disordered medium in a short sample length regime when the localization length ($\xi$) of the sample is much larger than the sample length (L), i.e., $L/\xi \to 0$ [17]. Using this correlation, it can be shown that real and imaginary delay times are also strongly correlated in this parameter regime of weak disorder and short sample length. In turn, the statistics of the real delay time, which is derived from the energy derivative of the real phase, and the imaginary delay time, which is derived in a similar way from the energy derivative of an 'imaginary phase', as obtained from reflection intensity/coefficient of a weakly disordered medium, have the same statistical form and a mirror image with a shift in time. Phase information is generally lost in the intensity expression as this is the square of the absolute value of the complex reflection amplitude. However, we show here, for the first time, that the delay time statistics for weakly disordered media can be recovered from the reflection statistics, thus eliminating the need for phase tracking, interferometry, or any other complex method for phase measurements in a weakly disordered medium. That is, the imaginary delay time, which is basically derived from the reflection coefficient, is the counter for the real delay time, which is derived from the phase, in one-dimensional systems.

The real delay time $\tau_r$, also often called sojourn time, is the time a wave spends in order to tunnel or travel through a potential barrier. In a random medium, delay time is a statistical quantity, similar to reflection coefficient, which provides a measure of the time spent by photons through multiple reflections before escaping from a sample. Consider the complex reflection amplitude (R) and the real delay time ($\tau_r$) represented as follows [3,15]:

$$R = r^{1/2} e^{i\theta(E)}, \tag{1}$$

$$\tau_r \equiv \frac{d\theta(E)}{dE} \propto \frac{d\theta(E)}{d(ck)} = \frac{d\theta(k)}{dk}. \tag{2}$$

In the above equations, we have considered the energy E of the wave/photons as $E = ck$, where k is the wave vector, and c is the velocity of light. To simplify the calculation, we will consider $c = 1$.

Likewise, one can derive the imaginary delay time from the magnitude of the reflection amplitude |R| by introducing an imaginary phase ($\theta_i$) defined as:

$$|R| = r^{1/2} = e^{-\theta_i} = e^{i(i\theta_i)} \tag{3}$$

$$\theta_i = -\ln(|R|) \tag{4}$$

$$\tau_i \equiv \frac{d\theta_i(E)}{dE} \propto \frac{d\theta_i(E)}{d(ck)} = \frac{d\theta(k)}{dk} = -\frac{1}{|R|}\frac{d|R|}{dk} \tag{5}$$

Consequently, an analogy between real and imaginary delay time can then be shown as

$$R = |R|e^{i\theta} = e^{-\theta_i}e^{i\theta} = e^{i(i\theta_i)}e^{i\theta}, \tag{6}$$

$$\theta_i = -\ln(|R|) \quad \text{since} \quad |R| = e^{-\theta_i}, \tag{7}$$

$$\therefore \tau_r = \tau = \frac{d\theta}{dk} \Leftrightarrow \tau_i = \frac{d\theta_i}{dk} \tag{8}$$

### 1.1. Delay time distributions $P(\tau_r)$ and $P(\tau_i)$ from the Langevin equation of the reflection coefficient

We numerically calculated real delay time, distribution $P(\tau_r)$, and imaginary delay time distribution $P(\tau_i)$ to evaluate any correlation resulting from the strong phase and amplitude correlations in the complex reflection amplitude (correlation calculation shown later). Consider a one-dimensional optical disordered medium with refractive index $n(x) = n_0 + dn(x)$ and length L and a light wave of wave vector k (i.e., $k = 2\pi/\lambda$, $\lambda$: wavelength of the light) reflected from the sample. In this case, a Langevin equation for the complex amplitude reflection coefficient R(L) can be derived as follows [16]:

$$\frac{dR(L)}{dL} = 2ikR(L) + \frac{ik}{2}\frac{2dn(L)}{n_0}[1 + R(L)]^2, \tag{9}$$

with the initial condition R(L=0) = 0. Now consider

$$R(L) = |R(L)|\exp(i\theta) = r(L)^{1/2}\exp(i\theta(L)). \tag{10}$$

Using the above relation for R(L), in the Langevin equation, we obtain two coupled differential equations in r and $\theta$ [16]:

$$\frac{dr(L)}{dL} = f_r(r,\theta) = k\frac{2dn}{n_0}r^{1/2}(1-r)\sin\theta, \quad (11)$$

$$\frac{d\theta(L)}{dL} = f_\theta(r,\theta) = 2k + \frac{k}{2}\frac{2dn}{n_0}[2+(r^{1/2}+r^{-1/2})\cos\theta]. \quad (12)$$

From Equations (11) and (12), the real delay time ($\tau_r$) and the imaginary delay time ($\tau_i$) can be derived as follows (with c=1):

$$\tau_r \equiv \frac{d\theta_r(E)}{dE} = \frac{d\theta_r(k)}{dE} = \frac{d}{dk}\int_0^L f_\theta(r,\theta)d\theta, \quad (13)$$

$$\tau_i \equiv \frac{d\theta_i(E)}{dE} = \frac{d\theta_i(k)}{dE} = \frac{1}{|R|}\frac{d}{dk}\int_0^L f_r(r,\theta)d\theta, \quad (14)$$

## 2. Numerical simulations and results for $\tau_i$ and $\tau_r$ statistics:

We performed numerical simulations for Equations (13) and (14) to calculate the values of $\tau_r$ and $\tau_i$ by direct stochastic simulation of R(L) in the visible light range. Simulations were performed for different sample lengths L, ranging from 1 - 40 μm over the visible light range between 500-700 nm. Keeping practical applications in mind, all studies were performed by considering refractive index fluctuation strength dn=0.02 and localization length $\xi$=140 μm, which are relevant to biological media, albeit in weakly disordered media and short length r=L/$\xi$ [15]. Furthermore, to better mimic a natural biological sample, we considered the Gaussian color noise model of refractive index fluctuations with short range exponential delay correlation of correlation length $l_c$~20 nm (<<L).

### 2.1. P($\tau_i$) and P($\tau_r$) statistics:

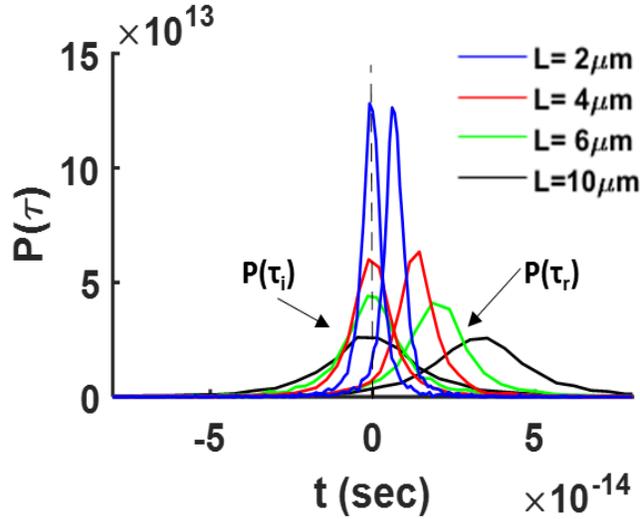

Fig.1. Probability distribution of real delay time P($\tau_r$) and distribution of imaginary delay time P($\tau_i$), for different sample lengths. Parameters: $n_0$=1, dn=0.02, $lc$=20nm, k=2π/λ (λ=600nm).

Figure 1 shows plots of the statistics for real delay time P($\tau_r$) and imaginary delay time P($\tau_i$), for different lengths. For a constant sample length L, it can be seen that the distribution of P($\tau_r$) has the same shape as that of P($\tau_i$), but a mirror image with a shift in time. While imaginary delay time is symmetrically distributed around mean zero, the average real delay time increases with the increase in sample length. Each distribution P($\tau_r$) is also symmetric around its respective mean value. In Figure 1, we see that the shape of two distributions at a fixed L appears to be same, mirror image with a shift. Therefore, in further analysis, different average values of delay time, as derived from the statistics of P($\tau_r$) and P($\tau_i$), such as mean real delay time $\langle\tau_r(L)\rangle$, mean imaginary delay time $\langle\tau_i(L)\rangle$, and their respective standard deviation $\sigma(\tau_r(L))$ and $\sigma(\tau_i(L))$, were analyzed as a function of sample length L. The results are shown in Figure 2.

A number of sample lengths ranging from 1 μm to 40 μm were studied, i.e., corresponding reflection coefficient in a range r~.15. Simulations were carried out with disorder parameter dn = 0.02 and at a wavelength λ = 0.6 μm, corresponding to k = 2π/λ = 10.47 μm$^{-1}$. Wavelength λ = 0.6 μm was chosen around the midpoint the visible spectrum (0.450 - 0.750 μm).

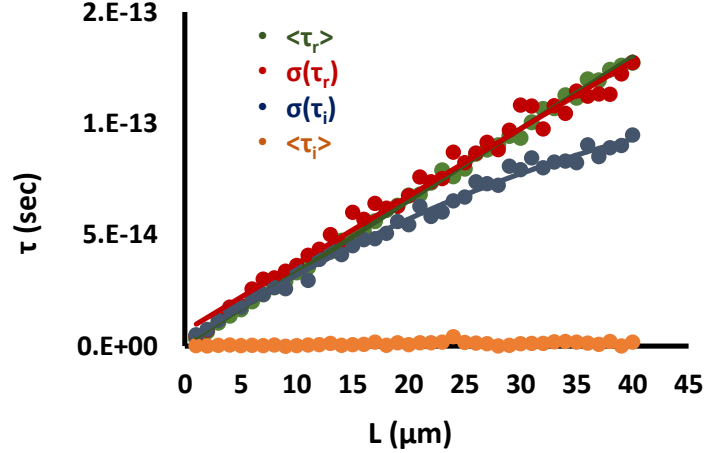

Fig. 2. Variation of the mean and *std* of real and imaginary delay time with length : Mean real delay time $\langle\tau_r\rangle$ , *std* of real delay time $\sigma(\tau_r)$, and mean imaginary delay time $\sigma(\tau_r)$, all are approximately same and overlapped in the variation with the length. However mean imaginary delay time $\langle\tau_i\rangle$=0, for all length due to the symmetry of P($\tau_i$) distribution around zero, as shown in Fig.1.

The results presented in Figure 2 show that $\langle\tau_r(L)\rangle$, $\sigma(\tau_r(L))$, and $\sigma(\tau_i(L))$ linearly increase with increase in the sample length from 1 μm to 40 μm. However, up to ~20 μm, the linearity of these three parameters are exactly the same, while $\langle\tau_i(L)\rangle$ remains zero for all whole sample length scales. It can be seen that the standard deviation of $\sigma(\tau_i(L))$ deviates with increasing length around sample length scale ~20 μm. This result is intriguing. To explain, at least three of the parameters, namely $\langle\tau_r(L)\rangle$, $\sigma(\tau_r(L))$, and $\sigma(\tau_i(L))$, show exactly the same linear dependence on sample length L < 20 μm. This suggests the existence and characterization of the delay time by only one parameter, in this sample length scale. Therefore, it is further suggested that this relationship can be exploited to reconstruct real delay time statistics from imaginary delay time based on the imaginary delay time statistics. In other words, since r ≃ L/ξ in this region, ξ~140μm and r~.14, the region can be generalized for any combination of L and ξ with the range up to r= L/ξ ~.14, and deviation will start after this value.

*2.2. Deriving real delay time information from imaginary delay time*

By understanding the statistics of imaginary time distribution, real time distribution can be constructed. That is, if imaginary time distribution is known, the value of real time distribution can be determined by simply shifting the mean of the distribution along the increasing direction by exactly the value of the standard deviation, as

$$P(\tau_r) \approx P(\tau_i + \sigma(\tau_i)). \qquad (15)$$

This relationship is valid for a range of sample length. Our work is largely driven by a weak disordered regime, one that is particularly relevant to biological systems, but it can be generalized for any disorder sample with range till r~.14. In this case, the valid range of the weak disorder is around ~ *kL* ~ (2π/.6)×20 ~ 210. This is a wide mesoscopic sample range.

3. **Strong correlation between phase and amplitude in a weakly disordered and short sample length:**

*3.1 Origin of strong correlation between r and θ*

Any complex reflection amplitude R [=$r^{1/2}\exp(i\theta)$] from a potential has two types of information: amplitude magnitude ($r^{1/2}$) and its phase (θ). For reflection from a one-dimensional delta reflection optical potential $n_0+n_\delta$ (air $n_0$=1) of width $l_w$, then the phase of the reflection coefficient can be shown as:

$$\theta = \tan^{-1}\left(\frac{2}{kl_w n_\delta}\right) \quad (16)$$

$$\Rightarrow \pm n_\delta \to 0, \quad \theta \approx \pm\pi/2 \quad . \quad (17)$$

This shows an approximate discrete values for phase, depending on the value of $n_\delta$. In particular, the phase and sign of the scattering potential, i.e., the effective two-state delta function random potential and its phase value, have strong correlation of occurrence at the phase value θ ≃ ±π/2. Thus, for a weakly disordered medium, this correlation brings the discrete focused value of increase in the reflection coefficient owing to the discrete values of the phase accumulation.

The differential equation for the reflection coefficient, i.e., Equation (11), can be now rewritten based on this correlation for the weak disorder limit and for a very small sample length where a strong correlation between the phase and amplitude exists, depending on the above correlation. Considering Equation (11) with L →0, the value of the reflection coefficient r( L) is a positive quantity and r (L)=0 at L =0 because the physical situation demands that dr/dL > 0 in the limit as L → +0 when sample length is very small. This condition means that dn.sin(θ) must be positive on the right-hand side of Equation (11) in this short length and weak disorder limit. In particular, sign(dn)=sign(sin(θ(L))), implying a strong correlation between phase and amplitude. This correlation is stronger for short-range correlation and weak disorder. The term "dn.sin(θ)" always indicates a positive quantity, which brings dn.sinθ to a positive definite magnitude |dn.sinθ|, a strongly correlated refractive index, and the phase along the positive direction for r ~0. Now, the reflection amplitude in Equation (11) can be written as

$$\frac{dr(L)}{dL} \approx \frac{2}{n_0} kr^{1/2}(1-r)|dn\sin\theta|. \quad (18)$$

*3.2 Simulation result for |R| and θ correlation*

A strong correlation can be seen between the phase and the sign of delta function strength with short spatial correlation length, essentially because dn and the phase angle θ~π/2 have the same sign. To determine the effect of this correlation on amplitude and phase in Equation (6), 1) for a finite length of the sample with Gaussian color noise disorder parameter, 2) in the biologically relevant parameter regime of refractive index (dn ~.02 ), and 3) with finite short sample length, we performed numerical simulation by direct integration of Equation (9). The simulations were conducted on four different sample lengths, namely 10, 20, 30 and 40 µm, with dn=0.02, lc =20nm, $n_0$=1, and wavelength λ=.6µm. The simulation results, as shown in Figure 4, show a remarkable correlation behavior between |R| and its phase θ. For smaller sample lengths, e.g., till ~20 µm, the phase angles were more discrete in nature by their concentration, either at -π/2 or π/2 for any value of |R|, demonstrating a strong correlation between |R| and θ. However, with increase in sample length, e.g., 30µm and 40µm, spread in distributions in phase, as well as in correlation, are obtained, suggesting decrease in the correlation between |R| and θ. The spread in θ and r correlation at higher sample lengths can be attributed to an increase of effective disorder-induced scattering from the sample resulting from the longer time spent by photons inside the sample, as well as more phase randomization and loss of correlation from stronger scattering. At higher sample lengths, the probability of obtaining a phase angle of +π/2 becomes more than that for obtaining θ=-π/2. Further increase in the length will show the uniform distribution of the phase P(θ) in this regime. This correlation is replicated in the expression of the delay time parameter regime, as described in Equation 2.

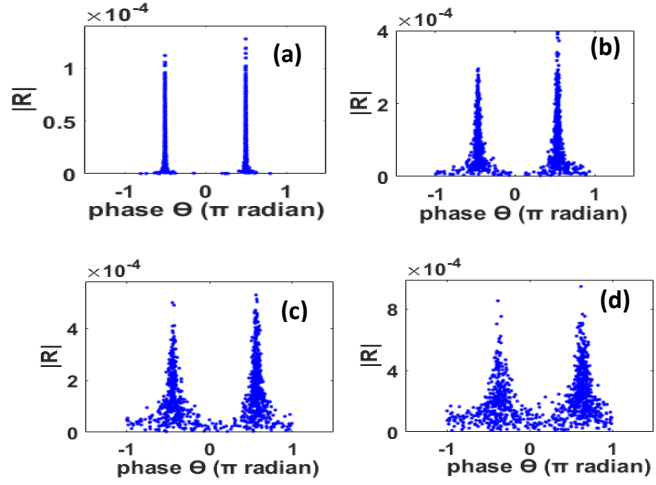

Fig. 3. Simulation study showing the correlation between |R| and θ conducted for samples (RI $n_0$=1 and fluctuation dn=0.02, $l_c$=20nm) of different lengths: (a) *L= 1.0 μm*, (b) *L= 10 μm*, (c) *L= 20 μm*, (d) *L=40 μm* with constant wavelength *λ= .6 μm* (around mid-visible spectrum). We can see a strong correlation between phase and amplitude in a weak disorder regime; however, this correlation decreases with the increase in sample length for a constant weak disorder dn=.02.

## 4. Asymptotic limit of infinite length verification of delay time distribution

Here we want to verify that our graphical reproduction of numerical simulation represents the asymptotic limit of large sample length. For a larger sample length and defining $\alpha=4\cdot(dn^2 k)^{-1}$, it was shown that delay time has an analytical form, as given below [10,11]:

$$P(\tau_r) = \frac{\alpha}{\tau_r^2} \exp\left(-\frac{\alpha}{\tau_r}\right). \tag{19}$$

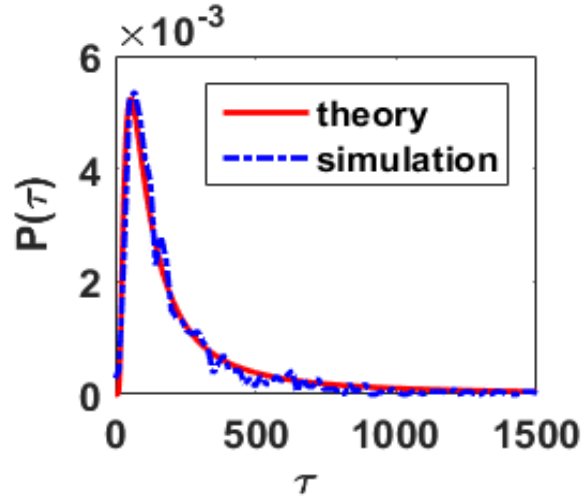

Fig.4. Verification that our numerical simulation can exactly replicate the known asymptotic limit of the equation.

We performed numerical simulations of real delay time distribution P($\tau_i$) with the asymptotic limit of the large length with weak disorder. In the large length limit, both numerical simulation form and analytical form match, confirming and validating, in turn, the algorithm used here for simulations.

## 5. Conclusion and Discussion

In this paper, we studied delay time statistics from one-dimensional weakly disordered optical samples. Real delay time refers to the extra time spent for a wave to tunnel or go through multiple reflections from a potential barrier or a disorder potential configuration. In random media, delay time is a statistical measure, just like reflectance, but its value is related to the derivative of the reflection amplitude, or the phase of the reflection coefficient. In a weak disorder regime, amplitude |R| and phase $\theta$ are strongly correlated in complex reflection amplitude ($R(L)=r(L)^{1/2}\exp(i\theta)$). In this regime, phase $\theta$ only has discrete values, depending on the net sign of the optical potential, in particular $\theta=\pm\pi/2$. This correlation can exist for a wide range of parameter space for a weak disorder regime and short sample length. From this correlation, in turn, it was further shown that the real delay time and corresponding imaginary delay time distributions have the same form/shape and mirror image with a shift in time. However, phase/amplitude correlation decreases with the increase of sample length and the strength of disorder. In turn, the real and imaginary delay time correlation also decreases with sample length and disorder strength. Many examples of weak disorder can be cited, varying from biological systems to polymers. For instance, biological samples, such as cells and tissues, are typical examples of weakly disordered media. As we have shown, in this regime of weak disorder and short length, reflection amplitude information is sufficient to derive the real delay time distribution, but without actually deriving the real phase and its derivative. Therefore, this technique will bring a new dimension to our understanding of phase statistics, as well as delay time, for weakly disordered media and samples of short length. The correlation is up to a range $r= L/\xi \sim .14$ (for any combinations of L and $\xi$). However, this correlation will be lost for greater lengths or the parameter regime when $\theta$ is uncorrelated, uniformly distributed, and independent of the r distribution. Under these conditions, real delay time cannot be derived from the imaginary delay time. Otherwise, since real delay time represents 1) the extra time it takes for a photon to tunnel through, or reflect from, the potential barrier and 2) the energy derivative of the phase of the reflection coefficient, it may be possible to utilize imaginary delay time statistics as a new marker, or parameter, to characterize biological cells ( or similar media), which are, inherently, weakly disordered media.

## Acknowledgements
The work was supported by the NIH, University of Memphis, and the FedEx Institute of Technology. We thank D. J. Park for useful discussions.